\documentclass[%
 reprint,
 amsmath,amssymb,
 aps,
]{revtex4-2}

\usepackage[T1]{fontenc} 
\usepackage{bm}
\usepackage{url}
\DeclareMathAlphabet{\pazocal}{OMS}{zplm}{m}{n}
\usepackage{braket}
\usepackage{amsmath}
\usepackage{amssymb}
\usepackage{leftidx}

\usepackage{bm}
\usepackage{slashed}
\usepackage{calrsfs}
\usepackage{cancel}
\usepackage{dirtytalk}
\usepackage{upgreek}
\usepackage{hyperref}
\usepackage{graphicx}
\usepackage{dcolumn}

\begin{document}

\preprint{APS/123-QED}

\title{Conservation of all Lipkin's zilches from symmetries of the standard electromagnetic action and a hidden algebra}

\author{Vasileios A. Letsios}
 \email{vasileios.letsios@york.ac.uk}
\affiliation{%
Department of Mathematics, University of York\\Heslington, York, YO10 5DD, United Kingdom 
}%

\begin{abstract} 
In 1964, Lipkin discovered the \textit{zilches}, a set of conserved quantities in free electromagnetism. Among the zilches, \textit{optical chirality} was identified by Tang and Cohen in 2010, serving as a measure of the handedness of light and leading to investigations into light's interactions with chiral matter. While the symmetries underlying the conservation of the zilches have been examined, the derivation of zilch conservation laws from symmetries of the standard free electromagnetic (EM) action using Noether's theorem has only been addressed in the case of optical chirality. We provide the full answer by demonstrating that the \textit{zilch symmetry transformations} of the four-potential, \texorpdfstring{$A_{\mu}$}{A}, preserve the standard free EM action. We also show that the zilch symmetries belong to the enveloping algebra of a "hidden" invariance algebra of free Maxwell's equations. This "hidden" algebra is generated by familiar conformal transformations and certain "hidden" symmetry transformations of~\texorpdfstring{$A_{\mu}$}{A}. Generalizations of the ``hidden'' symmetries are discussed in the presence of a material four-current, as well as in the theory of a complex Abelian gauge field. Additionally, we extend the zilch symmetries of the standard free EM action to the standard interacting action (with a non-dynamical four-current), allowing for a new derivation of the continuity equation for optical chirality in the presence of electric charges and currents. Furthermore, new continuity equations for the remaining zilches are derived.

\end{abstract}

\maketitle


\section{Introduction}

Noether's seminal theorem~\cite{Noether1918} is the cornerstone in understanding the deep connection between symmetries of physical theories and conservation laws. Starting from continuous symmetries of the action functional of a theory, Noether's theorem can be used to derive conservation laws for the associated Euler-Lagrange equations.
In relativistic field theories, such as electromagnetism in Minkowski spacetime, the knowledge of a symmetry leads to a Noether (four-)current, $V^{\mu}$, which is conserved ($\partial_{\mu}V^{\mu}=0$). This conservation holds for fields satisfying the Euler-Lagrange equations - i.e. for on-shell field configurations - and the corresponding Noether charge, $Q= \int d^{3}x V^{0}$, is time-independent. 

An example of little-known time-independent quantities in free electromagnetism is given by the ten \textit{zilches} that were discovered by Lipkin in 1964~\cite{Lipkin}. One of the zilches, now known as \textit{optical chirality}, started drawing renewed theoretical and experimental interest in 2010, when Tang and Cohen realized that this particular zilch provides a measure of the chirality (or handedness) of light~\cite{optical}. 
The optical chirality density for the free electromagnetic (EM) field is~\cite{Lipkin, optical}
\begin{align} 
    C=&~ \frac{1}{2}  \left( -\bm{E} \cdot \frac{\partial \bm{B}}{\partial t}  + \bm{B} \cdot\frac{\partial \bm{E}}{\partial t}  \right) \label{optical chirality density},
\end{align}
where $\bm{E}$ and $\bm{B}$ are the electric and magnetic fields, respectively~\footnote{An alternative expression for optical chirality density that appears in the literature is $C=~\tfrac{1}{2}  \left( \bm{E} \cdot \bm{\nabla} \times \bm{E}  + \bm{B} \cdot  \bm{\nabla} \times \bm{B} \right)$~\cite{optical}. This expression is equal to Eq.~(\ref{optical chirality density}) only in the absence of electric charges and currents. In this Letter, we use the expression~(\ref{optical chirality density}) when electric charges and currents are present. The justification for our choice is that the expression~(\ref{optical chirality density}) is equal to the $000$-component of the zilch tensor~(\ref{zilch_tensor}) (up to a factor of 1/2), while, as we show in this Letter, the zilch tensor is the Noether current corresponding to the zilches in free electromagnetism. Also, arguments in favor of defining the optical chirality density in the presence of electric charges and currents using Eq.~(\ref{optical chirality density}) can be found in Ref.~\cite{chirality-hel-etc}.}. (Throughout this Letter, we adopt the system of units in which the speed of light and the permittivity of free space are $c=\varepsilon_{0}=1$.) The flux of optical chirality is given by the three-vector
\begin{align}\label{flux of optical chirality}
    \bm{S}= \frac{1}{2} \bm{E} \times \frac{\partial \bm{E}}{\partial t} + \frac{1}{2} \bm{B} \times \frac{\partial \bm{B}}{\partial t},
\end{align}
while the differential conservation law for optical chirality~\cite{Lipkin}
\begin{align}\label{conservation optical}
    \frac{\partial}{\partial t}C+ \bm{\nabla} \cdot \bm{S}= 0
\end{align}
is satisfied if $\bm{E}$ and $\bm{B}$ obey the free Maxwell equations
\begin{align}\label{Maxwell_eq_usual}
&\bm{\nabla} \times \bm{B}=  \frac{\partial   \bm{E}}{\partial t},\hspace{4mm}\bm{\nabla} \times \bm{E}= - \frac{\partial   \bm{B}}{\partial t},\nonumber \\
  & \bm{\nabla}\cdot \bm{E} =0, \hspace{10mm}\bm{\nabla}\cdot \bm{B} =0.
\end{align}
Optical chirality is given by the integral of $C$ over the space, $\int d^{3}x \,C$, and is a constant of motion for free electromagnetism~\cite{Lipkin}.

 In Ref.~\cite{optical}, Tang and Cohen demonstrated that, in the presence of an EM field, the dissymmetry in the excitation rate of two small chiral molecules that are related to each other by mirror reflection is determined by the optical chirality. These findings have motivated novel investigations into chiral light-matter interactions~\cite{optical, TangCohenagain, Nanotech, Cohenenhanced, generation_of_chiralEM, ChiralPlasmNano, EnhancedOpticalChir,Poulikakos3Dprl, PoulikakosFluorDichroism, PoulikakosFlux, Poulikakosopticalantenna, Poulikakosproteins, PoulikakosPhD}.
Understanding these interactions is very important in various disciplines. For example, it is known that deriving products of a given handedness in chemical reactions can be crucial - because molecules of a given handedness must be used in order to design drugs without negative side-effects~\cite{thalidomide} - and chiral light has been suggested to serve as a useful tool in order to achieve this~\cite{homochiral, Enantioselective_homochiral, assymetric_catal_book}. Applications of chiral light to the detection and characterization of chiral biomolecules have been also discussed~\cite{Nanotech}.
As for the other nine zilches, recently, Smith and Strange shed light on the mystery of their physical meaning for certain topologically non-trivial vacuum EM fields~\cite{Smith_2018}.

Although the \textit{zilch symmetries} - i.e. the symmetries underlying the zilch conservation laws - and their generalization have been discussed in previous works~\cite{Ibragimov,Cameron_2012,zilchrev,SimulikI,SimulikII,Poland, Sudbery, Calcin, Anco, simulikPhysLet}, there are still certain gaps concerning our mathematical understanding of them. Most importantly, there is a gap in the literature concerning the explicit derivation of all zilch conservation laws from symmetries of the standard free EM action using Noether's theorem. In this Letter, we fill this gap and we also provide new insight concerning the zilch symmetries. Before proceeding to the main part of this article, let us discuss what is already known concerning the zilch symmetries in Subsection~\ref{Subsec_what_is_known}, as well as review the main findings of the present article in Subsection~\ref{Subsec_a_gap_main_aim}. For later convenience, we present here our notation and conventions.

\noindent \textit{\textbf{Conventions.}}\textemdash Greek tensor indices run from $0$ to $3$ and Latin tensor indices from $1$ to $3$. We follow the Einstein summation convention, while indices are raised and lowered with the mostly plus Minkowski metric $\eta_{\mu \nu}=\text{diag}(-1,1,1,1)$. A spacetime point in standard Minkowski coordinates is $x^{\mu}=(x^{0},x^{1},x^{2},x^{3})\equiv(t,x^{i})$. The totally antisymmetric tensors in 4 and 3 dimensions are $\epsilon^{\mu \nu \rho \sigma}$ and $\epsilon^{ijk}$, respectively ($\epsilon^{0123}=-\epsilon^{123}=-1$).



Let $A_{\mu}=(-\phi, \bm{A})$ denote the EM four-potential. The standard free EM action
\begin{align}\label{action_electric_magnetic}
    & S=\frac{1}{2}\int d^{4}x \left(\bm{E}\cdot \bm{E}- \bm{B}\cdot \bm{B}   \right), \nonumber\\
    \text{with}&\hspace{4mm} \bm{E}=-\frac{\partial  \bm{A} }{\partial t}-\bm{\nabla}\phi,\hspace{4mm}\bm{B}= \nabla \times \bm{A},
\end{align}
is expressed as
\begin{align}\label{action}
    S=&-\frac{1}{4} \int d^{4}x \, F^{\mu \nu} F_{\mu \nu}  ,  
\end{align}
where the antisymmetric EM tensor is defined as $F_{\mu \nu} \equiv \partial_{\mu}A_{\nu} - \partial_{\nu}A_{\mu}$ (with $F_{0i}=-E_{i}$ and $F_{ik}=\epsilon_{ikm}B^{m}$). We denote the dual EM tensor as $^{\star} F_{\mu \nu}= \frac{1}{2}\epsilon_{\mu \nu \rho \sigma} F^{\rho \sigma}$. The free Maxwell's equations $\partial^{\nu} F_{\nu \mu}=0$ are expressed in potential form as
\begin{align}\label{Maxwell eq potential}
   \Box A_{\mu}- \partial_{\mu}\partial^{\nu}A_{\nu}=0,
\end{align}
where $\Box = \partial^{\nu}   \partial_{\nu}$.
Because of the definition of $F_{\mu \nu}$ in terms of the four-potential, the equation
\begin{align}\label{Bianchi_usual}
  \partial_{\rho}F_{\mu \nu} + \partial_{\nu}F_{\rho \mu }+ \partial_{\mu}F_{\nu \rho}=0  
\end{align}
is identically satisfied. Equation~(\ref{Maxwell eq potential}), as well as the action~(\ref{action}), are invariant under infinitesimal gauge-transformations 
\begin{align}\label{gauge trans}
    \delta^{gauge}A_{\mu} = \partial_{\mu}a,
\end{align}
where $a$ is an arbitrary scalar function.


\subsection{What is known about the zilch symmetries?}\label{Subsec_what_is_known}

 
The zilch conservation laws can be conveniently described in terms of the zilch tensor~\cite{Lipkin, Kibble}
\begin{align}\label{zilch_tensor}
Z^{\mu}_{\hspace{2mm}\nu\rho}=&~-\,^{\star} F^{\mu \lambda}  \partial_{\rho}F_{ \lambda  \nu} + F^{\mu \lambda}~\partial_{\rho}\,^{\star} F_{ \lambda  \nu}.
    \end{align}
This is conserved on-shell, $\partial^{\rho} Z^{\mu}_{\hspace{2mm}\nu\rho} =0$, and the ten time-independent quantities~\cite{Lipkin}: $$\mathcal{Z}^{\mu \nu}= \mathcal{Z}^{\nu \mu} = \int d^{3}x Z^{\mu \nu 0}$$ are the ten zilches (see Section~\ref{Sec_zilch_tensor} for background material concerning the zilches). The optical chirality density~(\ref{optical chirality density}) is related to the zilch tensor as $Z^{000}= 2 C$.

At the level of free Maxwell's equations expressed in terms of the EM tensor, the zilch symmetries are known~\cite{SimulikI, Anco}. More specifically, the \textit{zilch symmetry transformations} of the EM tensor are~\cite{SimulikI, Anco} 
\begin{align}\label{transform_zilch_vecpot_FIELDSTRENGTH_INTRO}
  \Delta F_{\mu \nu}  &=~\tilde{n}^{\alpha}n^{\rho}\, \partial_{\alpha}\partial_{\rho}\hspace{0.05mm}^{\star}F_{\mu \nu},
\end{align}
where $\tilde{n}^{\alpha}$ and $n^{\rho}$ are two arbitrary constant four-vectors.
These transformations are symmetries of free Maxwell's equations, i.e. if $F_{\mu \nu}$ is a solution, then so is $\Delta F_{\mu \nu}$. In Ref.~\cite{Anco}, a complete classification of all independent local conservation
laws of Maxwell’s equations was given by using the methods described in Refs.~\cite{Olver_book, Anco_direct_constr}. Using these methods, it was shown that the zilch symmetries~(\ref{transform_zilch_vecpot_FIELDSTRENGTH_INTRO}) of free Maxwell's equations give rise to the conservation of the zilch tensor~(\ref{zilch_tensor}). However, in Ref.~\cite{Anco} the invariance of the standard EM action~(\ref{action}) was not discussed.

The zilch symmetries have also been studied in the case of duality-symmetric electromagnetism~\cite{zilchrev}. The duality-symmetric EM action is~\cite{Cameron_2012}
\begin{align}\label{dual-symmetric_action}
    \tilde{S}= -\frac{1}{8} \int d^{4}x \left( F^{\mu \nu}F_{\mu \nu}+ G^{\mu \nu} G_{\mu \nu}  \right).
\end{align}
This theory is an extension of the standard EM theory as it has two four-potentials, $A_{\mu}$ and $C_{\mu}$, and two EM tensors $F_{\mu \nu}= \partial_{\mu} A_{\nu}   - \partial_{\nu} A_{\mu}$ and $G_{\mu \nu}= \partial_{\mu} C_{\nu}   - \partial_{\nu} C_{\mu}$. The duality-symmetric theory coincides with the standard EM theory only after we impose the duality constraint $G_{\mu \nu}= \,^{*}F_{\mu \nu}$. 
In Ref.~\cite{zilchrev}, following the reverse Noether procedure, it was shown that the `generalized' version of the zilch tensor: 
\begin{align}\label{zilch_dualEM}
\mathcal{Z}^{\mu}_{\hspace{2mm}\nu\rho}=&~-\frac{1}{2}\,G^{\mu \lambda}  \partial_{\rho}F_{ \lambda  \nu} + \frac{1}{2} F^{\mu \lambda}~\partial_{\rho}G_{ \lambda  \nu} \nonumber \\
 &~-\frac{1}{2}G_{\nu}^{\hspace{1mm} \lambda}  \,\partial_{\rho}F_{ \lambda }^{\hspace{1.5mm}\mu}+ \frac{1}{2} F_{\nu}^{\hspace{1.5mm}\lambda}~\partial_{\rho}G_{ \lambda}^{\hspace{1mm}\mu}  
    \end{align}
is the Noether current corresponding to the following zilch symmetry transformations~\cite{zilchrev}:
\begin{align}\label{transform_zilch_vecpotentials_DUALEM}
    \tilde{\Delta} A_{\nu}&=~n^{\rho}\tilde{n}^{\mu}~\partial_{\rho}G_{\mu \nu}  \nonumber  \\
\tilde{\Delta} C_{\nu}&=-n^{\rho}\tilde{n}^{\mu}~\partial_{\rho}F_{\mu \nu}.
\end{align}
It has been shown that these transformations leave invariant the duality-symmetric action~(\ref{dual-symmetric_action})~\cite{zilchrev}. Then, the conservation of the zilches follows from the fact that the tensor~(\ref{zilch_dualEM}) coincides with the zilch tensor~(\ref{zilch_tensor_again}) of the standard EM theory if we apply the duality constraint. 

The derivation of the zilch conservation laws from symmetries of alternative actions has been studied in Refs.~\cite{Sudbery, SimulikI}.

\subsection{{Filling a gap in the literature, main results of this article and outline}} \label{Subsec_a_gap_main_aim}
In order to derive all zilch conservation laws using Noether's theorem in the case of standard electromagnetism, one needs to find the zilch symmetry transformations of the four-potential that leave the standard action~(\ref{action}) invariant. It is easy to observe that the zilch symmetry transformations $\Delta F_{\mu \nu}$~[Eq.~(\ref{transform_zilch_vecpot_FIELDSTRENGTH_INTRO})] of free Maxwell's equations are induced by the following zilch transformations of the four-potential:
\begin{align}\label{transform_zilch_vecpot_INTRO}
    \Delta A_{\nu} =~ n^{\rho}\tilde{n}^{\mu}\,  \epsilon_{\mu \nu \sigma  \lambda} \,\partial^{\sigma}\partial_{\rho}A^{\lambda}=n^{\rho}\tilde{n}^{\mu}~\partial_{\rho}\hspace{0.5mm}^{*}F_{\mu \nu} ,
\end{align}
with $\Delta F_{\mu \nu} \equiv \partial_{\mu} \Delta A_{\nu}- \partial_{\nu} \Delta A_{\mu}$ for on-shell field configurations. (The transformations~(\ref{transform_zilch_vecpot_INTRO}) coincide with $\tilde{\Delta} A_{\nu}$ in Eq.~(\ref{transform_zilch_vecpotentials_DUALEM}) if we apply the duality constraint.)~Interestingly, the study of the variation of the standard action~(\ref{action}) under the zilch transformations~(\ref{transform_zilch_vecpot_INTRO}) has not been studied in the literature. This means that the following question is still open:
\begin{align*}
    &\textit{How can we derive all zilch conservation laws from } \\ 
    &\textit{symmetries of the standard free EM action using } \\
    & \textit{Noether's theorem?}
\end{align*}
In this Letter, we give the full answer to this question by showing that the zilch transformations~(\ref{transform_zilch_vecpot_INTRO}) leave the standard EM action~(\ref{action}) invariant, and, then, we derive all zilch conservation laws using the standard Noether procedure~(see, e.g. Ref.~\cite{weinberg_1995}). 

Note that the only zilch conservation law that has hitherto been derived from symmetries of the standard action~(\ref{action_electric_magnetic}) is the one concerning the conservation of optical chirality~\cite{Philbin}.
 In particular, Philbin showed that optical chirality is the Noether charge corresponding to the following symmetry transformations~\cite{Philbin}:
 \begin{align}\label{Philbin vector form}
    \Delta \phi =0, \hspace{5mm} \Delta \bm{A}=  \bm{\nabla} \times \frac{\partial \bm{A}  }{\partial t}.
 \end{align}
This equation corresponds to a special case of the zilch symmetry transformation~(\ref{transform_zilch_vecpot_INTRO}) with $\tilde{n}^{\mu}=n^{\mu}= \delta^{\mu}_{0}$. In this article we provide an alternative (and covariant) derivation of Philbin's ~\cite{Philbin} result for optical chirality.

\noindent \textbf{Outline and main results.}~The basics concerning the zilch tensor and the zilches are reviewed in Section~\ref{Sec_zilch_tensor}. The derivation of all zilch conservation laws using the invariance of the standard action~(\ref{action}) under the zilch symmetries~(\ref{transform_zilch_vecpot_INTRO}) is presented in Section~\ref{Sec_action_Noether}. Then, we proceed by providing new insight concerning the conservation of the zilches and their underlying symmetries. More specifically, the rest of the investigations and findings of this article are summarized as follows:
 \begin{itemize}
     \item \textbf{A hidden invariance algebra of free Maxwell's equations and the zilch symmetries (Subsection~\ref{subsec_hidden_enveloping}).\textemdash}We show that the zilch symmetry transformations~(\ref{transform_zilch_vecpot_INTRO}) of the four-potential belong to the enveloping algebra of a ``hidden'' invariance algebra of free Maxwell's equations in potential form. This ``hidden'' algebra closes on the 30-dimensional real Lie algebra $so(6, \mathbb{C})_{\mathbb{R}}$ - i.e. the `realification' of the complex Lie algebra $so(6, \mathbb{C})$ - up to gauge transformations of the four-potential.~(The $so(6, \mathbb{C})_{\mathbb{R}}$ invariance of free Maxwell's equations in terms of the electric and magnetic fields was uncovered in Ref.~\cite{POHJANPELTO_letter}, but the potential form of Maxwell's equations was not discussed.)~The 30 generators of the ``hidden'' algebra correspond to the 15 well-known infinitesimal conformal transformations~[Eq.~(\ref{Lie deriv})] and to 15 little-known (``hidden'') infinitesimal transformations~[Eq.~(\ref{extra symmetry})]. The ``hidden'' transformations~(\ref{extra symmetry}) take a simpler form when acting on the EM tensor; that is a product of a duality transformation with an infinitesimal conformal transformation~\cite{SimulikI, SimulikII}~(see Eq.~(\ref{extra_symmetry_EM_tensor}))~\footnote{The fact that the product of a duality transformation
with an infinitesimal conformal transformation is a symmetry of Maxwell’s equations expressed in terms of the
EM tensor was first observed by Krivskii and Simulik~\cite{SimulikI, SimulikII}.}.

     \item \textbf{Hidden symmetries in the presence of matter (Subsection~\ref{subsect_hidden_matter}) and in the theory of a complex gauge field~(Subsection~\ref{Subsec_complex hidden}).\textemdash}We show that the ``hidden'' symmetries~[Eq.~(\ref{extra symmetry})] of free Maxwell's equations persist in the presence of a material four-current~[see Eq.~(\ref{extra symmetry matter})]. However, unlike the free case, the invariance algebra does not close on $so(6, \mathbb{C})_{\mathbb{R}}$. Then, we observe that the ``hidden'' symmetries of the real potential $A_{\mu}$ also exist for the free field equations~(\ref{Maxwell eq potential complex}) of a complex Abelian gauge field $\mathcal{A}_{\mu}$ - this is related to the complex formulation of duality-symmetric electromagnetism with the complex potential given by $\mathcal{A}_{\mu}= A_{\mu} + i C_{\mu}$~\cite{zilchrev}. We show that if we redefine the ``hidden'' transformations of $\mathcal{A}_{\mu}$ by multiplying with $i= \sqrt{-1}$, the 30-dimensional algebra becomes $so(4,2) \bigoplus so(4,2)$ (it closes again up to gauge transformations of the complex potential).

     \item \textbf{Zilch continuity equations from symmetries in the presence of matter~(Section~\ref{Section_continuity_inter}) and a new question~(Section~\ref{Sec_open_question}).\textemdash}We also study the derivation of zilch continuity equations in the presence of electric charges and currents by extending the zilch symmetries of the standard free action~(\ref{action}) to zilch symmetries [Eqs.~(\ref{transform_zilch_vecpot_AND_current_1})~and~(\ref{transform_zilch_vecpot_AND_current_2})] of the standard interacting action~(\ref{intercting_action}) (in which $A_{\mu}$ couples to a non-dynamical material four-current $J^{\mu}$). Taking advantage of the invariance of the interacting action under the zilch symmetries, we present a new way to derive the known continuity equation for optical chirality~\cite{ chirality-hel-etc}
\begin{align}\label{conservation optical_sources}
    \frac{\partial}{\partial t}C+ \bm{\nabla} \cdot \bm{S}= \frac{1}{2}\left( \bm{j}\cdot \frac{\partial \bm{B}}{\partial t}-\frac{\partial \bm{j}}{\partial t}\cdot {\bm{B}}\right)
\end{align}
($\bm{j}$ is the material electric current density). In Ref.~\cite{chirality-hel-etc}, the continuity equation~(\ref{conservation optical_sources}) was obtained from the complementary fields formalism, while a similar continuity equation had been first obtained in Ref.~\cite{optical}.
Apart from Eq.~(\ref{conservation optical_sources}), in this Letter, we also obtain new continuity equations [Eqs.~(\ref{cont_equation_charges_from_Noether})~and~(\ref{zilch_cont_charges})] for the rest of the zilches in the presence of electric charges and currents from symmetries of the interacting EM action~(\ref{intercting_action}). Then, we pose the interesting open question of whether the aforementioned invariance of the interacting EM action with a non-dynamical material four-current can be extended to the case where the material four-current is dynamical.

 \end{itemize}


\section{Background material concerning the zilch tensor and the zilches}\label{Sec_zilch_tensor}
In this Section we review the basics concerning the zilch tensor and the zilches.

The zilch tensor~(\ref{zilch_tensor}) can be expressed in various forms~\cite{Lipkin, Kibble}. For example, using the following identity~\cite{Kibble}:
   \begin{align}\label{Kibble's identity}
      \partial_{\rho}\left( ^{\star}F_{ \lambda \nu}F^{\mu \lambda}\right)=-\frac{1}{4}\delta^{\mu}_{\nu}\,\partial_{\rho}\left(\,^{\star}F^{\lambda \kappa}F_{\lambda \kappa}\right),
   \end{align}
   the zilch tensor~(\ref{zilch_tensor}) can be equivalently expressed as
    \begin{align}
  Z^{\mu}_{\hspace{2mm}\nu\rho} =&~ -\,^{\star}F^{\mu \lambda}  \partial_{\rho}F_{ \lambda  \nu}-\,^{\star}F_{\nu}^{\hspace{1mm} \lambda}  \partial_{\rho}F_{ \lambda}^{\hspace{1mm}\mu} - \frac{1}{2}\delta^{\mu}_{\nu}\,^{\star}F^{\lambda \kappa} \partial_{\rho}F_{\lambda  \kappa}.
\end{align}
This expression makes manifest that the properties $Z^{\mu \nu}_{\hspace{3mm}\rho} = Z^{\nu \mu}_{\hspace{3mm}\rho}$ and $Z^{\mu}_{\hspace{3mm}\mu \rho}=0$ are identically satisfied. Moreover, by using free Maxwell's equations, it is straightforward to show that the zilch tensor is divergence-free with respect to all of its indices and also satisfies $Z^{\rho}_{\hspace{3mm}\nu \rho}=0$~\cite{Kibble}. Using the fact that the zilch tensor is symmetric in its first two indices we can rewrite Eq.~(\ref{zilch_tensor}) as
\begin{align}\label{zilch_tensor_again}
Z^{\mu}_{\hspace{2mm}\nu\rho}=&~-\frac{1}{2}\,^{\star} F^{\mu \lambda}  \partial_{\rho}F_{ \lambda  \nu} + \frac{1}{2} F^{\mu \lambda}~\partial_{\rho}\,^{\star} F_{ \lambda  \nu} \nonumber \\
 &~-\frac{1}{2}\,^{\star} F_{\nu}^{\hspace{1mm} \lambda}  \partial_{\rho}F_{ \lambda }^{\hspace{1mm}\mu}+ \frac{1}{2} F_{\nu}^{\hspace{1mm}\lambda}~\partial_{\rho}\,^{\star} F_{ \lambda}^{\hspace{1mm}\mu}  .
    \end{align}
 
As mentioned in the Introduction, the ten zilches are given by the following ten time-independent quantities~\cite{Lipkin, Kibble}:
\begin{align}\label{def_the_zilches}
    \mathcal{Z}^{\mu \nu}
    =\mathcal{Z}^{\nu \mu}= \int d^{3}x Z^{\mu \nu 0  },
\end{align}
with $\partial \mathcal{Z}^{\mu \nu} / \partial{t}=0$.
 Only nine zilches in Eq.~(\ref{def_the_zilches}) are independent since $Z^{\mu}_{\hspace{2mm}\mu \,0}=0$. The $\mu \nu 0$-component ($Z^{\mu \nu 0}$) of the zilch tensor is the spatial density of the zilch $\mathcal{Z}^{\mu \nu}$, and the $\mu \nu j$-components ($Z^{\mu \nu j}$) are the components of the three-vector describing the corresponding flux~\cite{Lipkin}. The time-independence of the ten zilches follows from the ten differential conservation laws described by $\partial_{\rho} Z^{\mu \nu \rho}=0$. The conservation law~(\ref{conservation optical}) for optical chirality corresponds to
$ \tfrac{1}{2}\left(\partial_{0}Z^{0 0 0 } +\partial_{j}Z^{00 j}\right)=0$.

 For later convenience, note that the integral in Eq.~(\ref{def_the_zilches}) has the symmetry property
\begin{equation}\label{symmetry_zilches}
   \int d^{3}x Z^{\mu \nu 0  } =\int d^{3}x Z^{\mu 0 \nu}\, \left(=\int d^{3}x Z^{0 \mu \nu}\right) 
\end{equation}
because the difference 
$Z^{\mu  \nu 0 }-Z^{\mu 0 \nu }$ can always be expressed as a spatial divergence~\cite{Kibble} $$Z^{\mu  \nu 0 }-Z^{\mu 0 \nu }= \partial_{j}  \Lambda^{\mu \nu j},$$
where the explicit expression for the tensor $\Lambda$ is not needed for the present discussion~\footnote{The interested reader can find the expression for $\Lambda$ from equation~(14) of Ref.~\cite{Kibble} or they can set $J^{\mu}=0$ and let $\rho=0$ in Eq.~(\ref{Kibble's_Eq_14}) of the present article.}. It immediately follows that the difference $Z^{ \mu 0 \nu}-Z^{ \nu 0 \mu }$ can also be written as a spatial divergence. Hence, the $\mu \nu$-zilch, $\mathcal{Z}^{\mu \nu}$, can be actually interpreted as the time-independent quantity that corresponds to any of the three differential conservation laws: $\partial_{\rho}Z^{\mu \nu \rho}=0$ (which is the one used by Lipkin~\cite{Lipkin}), $\partial_{\rho}Z^{\mu  \rho  \nu}=0$ and $\partial_{\rho}Z^{\nu \rho \mu}=0$.
These differential conservation laws are not independent of each other. For example, the conservation law $\partial_{\rho}Z^{\mu \nu \rho}=0$ can be re-written as $\partial_{\rho}Z^{\mu  \rho  \nu}=0$ by using the relations
\begin{align}\label{relate_densities}
   \partial_{0}Z^{\mu \nu 0} = \partial_{0}\left(Z^{\mu 0 \nu} + \partial_{j} \Lambda^{\mu \nu j}    \right) 
\end{align}
and
\begin{align}\label{relate_fluxes}
     \partial_{j}Z^{\mu \nu j} = \partial_{j}\left(Z^{\mu j \nu} - \partial_{0} \Lambda^{\mu \nu j}    \right) 
\end{align}
for the corresponding spatial densities and fluxes, respectively. 

\section{Conservation laws for all zilches from the invariance of the standard action under the zilch transformations  }\label{Sec_action_Noether}

In this Section we show that the zilch symmetry transformation~(\ref{transform_zilch_vecpot_INTRO}), which is given here again for convenience:
\begin{align}\label{transform_zilch_vecpot}
    \Delta A_{\nu} =~ n^{\rho}\tilde{n}^{\mu}\,  \epsilon_{\mu \nu \sigma  \lambda} \,\partial^{\sigma}\partial_{\rho}A^{\lambda}=n^{\rho}\tilde{n}^{\mu}~\partial_{\rho}\hspace{0.5mm}^{*}F_{\mu \nu} ,
\end{align}
is a symmetry of the standard free EM action~(\ref{action}). Then, we derive all zilch conservation laws using Noether's theorem.

Let us start by examining the way in which the zilch symmetry transformation~(\ref{transform_zilch_vecpot}) acts on the EM tensor for off-shell field configurations; that is
\begin{align}\label{transform_zilch_vecpot_FIELDSTRENGTH}
  \Delta F_{\mu \nu} &\equiv ~ \partial_{\mu} \Delta A_{\nu}   -  \partial_{\nu} \Delta A_{\mu}  \nonumber \\
  &=~\tilde{n}^{\alpha}n^{\rho}\,\left( \partial_{\alpha}\partial_{\rho}\hspace{0.05mm}^{\star}F_{\mu \nu}-\epsilon_{\alpha \mu \nu \sigma}\partial_{\rho}\partial_{\lambda}F^{\lambda \sigma} \right),
\end{align}
where we have made use of the following important off-shell identity~\footnote{Equation~(\ref{Bianchi dual}) can be readily proved by contracting with $\epsilon_{\gamma \delta}^{\hspace{4mm}\mu \nu}$ and then using well-known properties of the totally antisymmetric tensor.}:
\begin{align}\label{Bianchi dual}
    \partial_{\alpha}\,^{\star}F_{\mu \nu}+\partial_{\nu}\,^{\star}F_{\alpha \mu}+\partial_{\mu}\,^{\star}F_{\nu \alpha} = \epsilon_{\alpha \mu \nu \sigma} \partial^{\beta}F^{\hspace{2mm}  \sigma}_{\beta}.
\end{align}

We now proceed to demonstrate that the zilch symmetry transformation~(\ref{transform_zilch_vecpot}) is indeed a symmetry of the action~(\ref{action}) and then apply Noether's theorem.
We find that the variation
\begin{align}\label{variation_free_action_before_start}
   \Delta S  = -\frac{1}{2}\int d^{4}x\, F^{\mu \nu}  \,\Delta F_{\mu \nu} 
\end{align}
is given by a total divergence (without making use of the equations of motion), as
\begin{align}\label{variation_free_action}
    \Delta S=\int d^{4}x\,\partial_{\nu} D^{\nu}
\end{align}
with
\begin{align} \label{Dnu}
    D^{\nu}=\frac{1}{2}n^{\rho} \tilde{n}^{\mu} \left(2\,^{\star}F^{\lambda \nu} \partial_{\rho} F_{\mu \lambda} +Z_{\mu \hspace{2mm}\rho}^{\hspace{1.5mm}\nu}  +\delta^{\nu}_{\rho}\,^{\star}F_{\mu \sigma}\,\partial^{\beta}F_{\beta}^{\hspace{1mm}\sigma} \right)
\end{align}
- see Appendix~\ref{appendix_invariance} for some details of the calculation.
Now, the usual procedure~\cite{weinberg_1995} can be followed in order to construct the conserved Noether current, $V^{\nu}$, associated with the zilch symmetry transformation~(\ref{transform_zilch_vecpot}), as
\begin{align}\label{Noether current_general_formula}
    V^{\nu }=&~\frac{\partial{\mathcal{L}}}{ \partial (\partial_{\nu}A_{\mu}) } \Delta A_{\mu} -D^{\nu}, \end{align}
where $\mathcal{L}= -\tfrac{1}{4}F^{\alpha \beta}F_{\alpha \beta}$ is the free EM Lagrangian density. Substituting the expression for $D^{\nu}$ [Eq.~(\ref{Dnu})] into Eq.~(\ref{Noether current_general_formula}) and making use of the identity~(\ref{Kibble's identity}), we find
    \begin{align}\label{Noether current zilch}
  V^{\nu }  =&~\frac{1}{2}n^{\rho} \tilde{n}^{\mu} \left(   Z_{\mu\hspace{2mm}\rho}^{\hspace{1.5mm}\nu} - \delta^{\nu}_{\rho}\,^{\star}F_{\mu \sigma}\,\partial^{\beta}F_{\beta}^{\hspace{1mm}\sigma}\right).
\end{align}
The definition of a conserved Noether current is not unique; we are free to add any term that vanishes on-shell and/or any term that is equal to the divergence of any rank-2 antisymmetric tensor to the expression for the Noether current~\cite{Ramond:1981pw}. Thus, we are allowed to express the Noether current in Eq.~(\ref{Noether current zilch}) as
\begin{align}\label{Noether current zilch final}
    V^{\nu }_{zilch}=&~\frac{1}{2}n^{\rho} \tilde{n}^{\mu}Z_{\mu\hspace{2mm}\rho}^{\hspace{1.5mm}\nu} 
\end{align}
with $\partial_{\nu} V_{zilch}^{\nu}=0$.
 Since the constant four-vectors $n^{\rho}$ and $\tilde{n}^{\mu}$ in Eq.~(\ref{Noether current zilch final}) are arbitrary, we conclude that 
 \begin{align}
     \partial_{\nu} Z^{\mu \nu \rho} = 0. 
 \end{align}
In other words, the zilch tensor is the conserved Noether current corresponding to the zilch symmetries~(\ref{transform_zilch_vecpot}) of the standard free action~(\ref{action}), while the corresponding Noether charges are the zilches~(\ref{def_the_zilches}). 


\section{``Hidden'' symmetries}

\subsection{``Hidden'' invariance algebra of free Maxwell's equations and the zilch symmetries}\label{subsec_hidden_enveloping}
Here we investigate the relation of the zilch symmetry transformations~(\ref{transform_zilch_vecpot}) to a ``hidden'' $so(6,\mathbb{C})_{\mathbb{R}}$ invariance algebra of free Maxwell's equations in potential form~(\ref{Maxwell eq potential}). 

Let $\xi^{\mu}$ denote any of the fifteen conformal Killing vectors of Minkowski spacetime satisfying
\begin{align}\label{conf Killing eq}
    \partial_{\mu} \xi_{\nu}+ \partial_{\nu}  \xi_{\mu}= \frac{\partial^{\alpha}  \xi_{\alpha}}{2}\eta_{\mu \nu}.
\end{align}
The conformal Killing vectors $\xi^{\mu}$ of Minkowski spacetime consist of~\cite{CFTbook}: the four generators of spacetime translations,
\begin{align}\label{transl Killing}
    P_{(\alpha)} = P^{\mu}_{(\alpha)} \partial_{\mu}= \partial_{\alpha} ,
\end{align}
the six generators of the Lorentz algebra $so(3,1)$, 
\begin{align}\label{Lorentz Killing}
 M_{( \beta, \gamma)}= M_{( \beta, \gamma)}^{\mu} \partial_{\mu}=  x_{\beta}  \partial_{\gamma} - x_{\gamma}  \partial_{\beta},
\end{align}
the generator of dilations
\begin{align}\label{dilation}
   D=D^{\mu}\partial_{\mu}=x^{\mu} \partial_{\mu},
\end{align}
and the four generators of special conformal transformations
\begin{align}\label{special conformal}
   K_{(\alpha)}= K_{(\alpha)}^{\mu} \partial_{\mu} =x^{\nu}x_{\nu} \partial_{\alpha}-2 x_{\alpha} x^{\mu} \partial_{\mu}.
\end{align}
These fifteen vectors form a basis for the algebra of infinitesimal conformal transformations of Minkowski spacetime which is isomorphic to $so(4,2)$.

The ``hidden'' invariance algebra of free Maxwell's equations~(\ref{Maxwell eq potential}) is generated by two types of infinitesimal symmetry transformations of the four-potential. The first type corresponds to the well-known infinitesimal conformal transformations, conveniently described by the Lie derivative
\begin{align}\label{Lie deriv}
    L_{\xi}A_{\mu}= \xi^{\lambda}  \partial_{\lambda}A_{\mu}+ A_{\lambda} \partial_{\mu}\xi^{\lambda},\hspace{5mm} \xi \in so(4,2).
\end{align}
These transformations generate a representation of $so(4,2)$ on the solution space of Maxwell's equations~(\ref{Maxwell eq potential}).
The second type of transformations corresponds to the little-known (``hidden'') transformations~\cite{Pohjanpelto_book}
 \begin{align}\label{extra symmetry}
    T_{\xi}A_{\mu}=&\,~ \xi^{\rho}\epsilon_{\rho \mu \sigma \lambda } \partial^{\sigma} A^{\lambda},\hspace{5mm} \xi \in so(4,2).
  \end{align}
If $A_{\mu}$ is a solution of Maxwell's equations, then so are $L_{\xi} A_{\mu}$ and $T_{\xi}A_{\mu}$ for all $\xi \in so(4,2)$~\cite{Pohjanpelto_book}. The effect of the ``hidden'' transformation~(\ref{extra symmetry}) on $F_{\mu \nu}$ corresponds to the product of a duality transformation with an infinitesimal conformal transformation as
\begin{align}\label{extra_symmetry_EM_tensor}
    T_{\xi}F_{\mu \nu} \equiv ~ \partial_{\mu}T_{\xi}A_{\nu} - \partial_{\nu}T_{\xi}A_{\mu}  =&~ L_{\xi}\,^{\star}F_{\mu \nu},
\end{align}
where
\begin{align}\label{Lie_dual_field_strength}
    L_{\xi}\,^{\star}F_{\mu \nu} = \xi^{\rho}\partial_{\rho}\,^{\star}F_{\mu \nu} + \,^{\star}F_{\rho \nu}\,\partial_{\mu}\xi^{\rho} +\,^{\star}F_{\mu\rho }\,\partial_{\nu}\xi^{\rho}.
\end{align}
This symmetry transformation of the EM tensor was first found in Refs.~\cite{SimulikI, SimulikII}.

The structure of the ``hidden'' invariance algebra of Maxwell's equations in potential form is determined by the Lie brackets:
\begin{align}\label{[Lie,Lie]}
    [L_{\xi'},L_{\xi}]A_{\mu} =~L_{[\xi',\xi]}A_{\mu}
\end{align}
\begin{align}\label{Lie,extra}
    [{L}_{\xi'}, T_{\xi}]A_{\mu} =~ T_{[\xi',\xi]}A_{\mu}
\end{align}
and
\begin{align}\label{[extra,extra]}
    [T_{\xi'}, T_{\xi}]A_{\mu}=&-{L}_{[\xi',\xi]}A_{\mu} 
    +\partial_{\mu}\left([\xi',\xi]^{\sigma}A_{\sigma}-\xi'_{\sigma}\xi_{\rho}F^{\sigma \rho}   \right),
\end{align}
where, e.g., $[L_{\xi'},L_{\xi}]=L_{\xi'}L_{\xi} - L_{\xi}L_{\xi'}$, while $\xi$ and $\xi'$ are any two basis elements of $so(4,2)$ with $[\xi',\xi ]^{\rho}= L_{\xi'}\xi^{\rho}$.
 We observe the appearance of a gauge transformation of the form~(\ref{gauge trans}) in the last term of Eq.~(\ref{[extra,extra]}).
 To the best of our knowledge, the explicit expressions for the commutators~(\ref{Lie,extra}) and (\ref{[extra,extra]}) appear here for the first time.
The commutation relations in Eqs.~(\ref{[Lie,Lie]})-(\ref{[extra,extra]}) coincide with the commutation relations of the 30-dimensional real Lie algebra $so(6,\mathbb{C})_{\mathbb{R}}$~\cite{POHJANPELTO_letter} (up to the gauge transformation in Eq.~(\ref{[extra,extra]})).

 Now, let us denote the zilch symmetry transformation~(\ref{transform_zilch_vecpot}) with associated Noether current corresponding to $Z_{\alpha\hspace{2mm}\beta}^{\hspace{1.5mm}\nu}$ ($\alpha$ and $\beta$ have fixed values) as $\Delta_{(\beta, \alpha)}A_{\mu}$. The latter is readily expressed as [see Eq.~(\ref{transform_zilch_vecpot})]
\begin{align}\label{transform_zilch_vecpot_alpha_beta}
    \Delta_{(\beta, \alpha)} A_{\mu} = \partial_{\beta}\left(\epsilon_{\alpha \mu \sigma  \lambda} \,\partial^{\sigma}A^{\lambda}\right) = L_{P_{(\beta)}} T_{P_{(\alpha)}}A_{\mu} .
\end{align}
It is obvious from this expression that $\Delta_{(\beta, \alpha)} A_{\mu}$ is given by the product of a ``hidden'' transformation~(\ref{extra symmetry}) with respect to the translation Killing vector $P_{(\alpha)}=\partial_{\alpha}$ and a Lie derivative~(\ref{Lie deriv}) with respect to the translation Killing vector $P_{(\beta)}= \partial_{\beta}$. This makes clear that the zilch symmetry transformation $\Delta_{(\beta, \alpha)} A_{\mu}$ belongs to the enveloping algebra of our ``hidden'' invariance algebra [and so do all transformations of the form~(\ref{transform_zilch_vecpot})].

\subsection{``Hidden'' symmetries of Maxwell's equations in the presence of a material four-current}\label{subsect_hidden_matter}
In the presence of a material four-current Maxwell's equations are 
\begin{align}\label{Maxwell eq potential matter}
   \Box A_{\mu}- \partial_{\mu}\partial^{\nu}A_{\nu}=-J_{\mu},
\end{align}
where $J^{\mu}=(\rho,\bm{j})$ and $\partial_{\mu} J^{\mu}=0$. Maxwell's equations remain invariant under simultaneous infinitesimal conformal transformations of $A_{\mu}$ and $J_{\mu}$, i.e. Eq.~(\ref{Maxwell eq potential matter}) will still be satisfied if we make the following replacements:
\begin{align}\label{replacements_conformal_matter}
    &A_{\mu}\rightarrow L_{\xi}A_{\mu}, \nonumber \\ 
   &J_{\mu}\rightarrow L_{\xi}J_{\mu}+\frac{\partial_{\alpha} \xi^{\alpha}}{2} J_{\mu} ,\hspace{4mm} \xi \in so(4,2),
\end{align}
where $L_{\xi}$ is the Lie derivative~(\ref{Lie deriv}).
It is interesting to investigate whether the ``hidden'' symmetries~(\ref{extra symmetry}) of free Maxwell's equations also persist in the presence of matter. Indeed, we find that if $A_{\mu}$ and $J_{\mu}$ satisfy Eq.~(\ref{Maxwell eq potential matter}), then Eq.~(\ref{Maxwell eq potential matter}) will still be satisfied if we make the following replacements:
\begin{align}\label{extra symmetry matter}
 &A_{\mu}\rightarrow   T_{\xi}A_{\mu},\nonumber\\
 &J_{\mu}\rightarrow  \delta^{hid}_{\xi}J_{\mu}= \epsilon_{\rho \mu \sigma \lambda } \partial^{\sigma}\left( \xi^{\rho}J^{\lambda} \right),
 \hspace{5mm} \xi \in so(4,2),
  \end{align}
where $T_{\xi}A_{\mu}$ is given by Eq.~(\ref{extra symmetry}), while we call $\delta^{hid}_{\xi}J_{\mu}$ in the second line the ``hidden'' transformation of the four-current. Equation~(\ref{extra symmetry matter}) describes the ``hidden'' symmetries of Maxwell's equations in the presence of a material four-current.

Unlike the free case, in the presence of matter, the algebra does not close on $so(6,\mathbb{C})_{\mathbb{R}}$ up to gauge transformations of the four-potential~\footnote{If one considers only the familiar conformal transformations~(\ref{replacements_conformal_matter}), then the algebra closes on $so(4,2)$.}. This can be readily understood from the following example. By calculating the commutator between ``hidden'' symmetries generated by translation Killing vectors~(\ref{transl Killing}), we find:
\begin{align}\label{[extratrans,extratrans]A matter }
    &[T_{P_{(\alpha)}}, T_{P_{(\beta)}}]A_{\mu}\nonumber\\
    =&~\partial_{\mu}\left(-P_{(\alpha)}^{{\sigma}}P^{\rho}_{(\beta)}F_{\sigma \rho}   \right) +\left( P_{(\alpha) \mu}P^{\rho}_{(\beta)}  -P^{\rho}_{(\alpha) }P_{(\beta) \mu} \right)  \partial^{\sigma} F_{\sigma \rho}
\end{align}
(compare this equation with Eq.~(\ref{[extra,extra]})). In the absence of matter, the second term in Eq.~(\ref{[extratrans,extratrans]A matter }) vanishes and the algebra closes up to the gauge transformation $\partial_{\mu}\left(-P_{(\alpha)}^{{\sigma}}P^{\rho}_{(\beta)}F_{\sigma \rho}   \right)$ - see Eq.~(\ref{[extra,extra]}). However, in the presence of matter, the second term in Eq.~(\ref{[extratrans,extratrans]A matter }) seems to describe a new symmetry transformation. Similarly, we find the commutator for the four-current:
\begin{align}\label{[extratrans,extratrans]J matter }
    [\delta^{hid}_{P_{(\alpha)}}, \delta^{hid}_{P_{(\beta)}}]J_{\mu}
    =&~\partial_{\mu}\left(-P_{(\alpha)}^{{\sigma}}P^{\rho}_{(\beta)}(\partial_{\sigma}J_{\rho} -\partial_{\rho}J_{\sigma} )   \right) \nonumber\\ &+\left( P_{(\alpha) \mu}P^{\rho}_{(\beta)}  -P^{\rho}_{(\alpha) }P_{(\beta) \mu} \right)  \Box J_{\rho}.
\end{align}
From Eqs.~(\ref{[extratrans,extratrans]A matter }) and (\ref{[extratrans,extratrans]J matter }), it follows that Maxwell's equations~(\ref{Maxwell eq potential matter}) will still be satisfied~(this is easy to verify) if we simultaneously make the replacements:
\begin{align}\label{weird_replacements}
    &A_{\mu}\rightarrow \left( P_{(\alpha) \mu}P^{\rho}_{(\beta)}  -P^{\rho}_{(\alpha) }P_{(\beta) \mu} \right)  \partial^{\sigma} F_{\sigma \rho}
    \end{align} 
    and
   \begin{align}
   J_{\mu}\rightarrow &~\partial_{\mu}\left(-P_{(\alpha)}^{{\sigma}}P^{\rho}_{(\beta)}(\partial_{\sigma}J_{\rho} -\partial_{\rho}J_{\sigma} )   \right) \nonumber\\ &+\left( P_{(\alpha) \mu}P^{\rho}_{(\beta)}  -P^{\rho}_{(\alpha) }P_{(\beta) \mu} \right)  \Box J_{\rho}.
\end{align}
The study of the full structure of the algebra in the presence of matter is something that we leave for future work.

\subsection{``Hidden'' symmetries for the complex Abelian gauge field}\label{Subsec_complex hidden}
The free (hermitian) action for the complex Abelian gauge field, $\mathcal{A}_{\mu}$, is
\begin{align}\label{dual-symmetric_action complex}
     -\frac{1}{8} \int d^{4}x  \mathcal{F}^{\dagger}_{ \mu \nu}\mathcal{F}^{\mu \nu},
\end{align}
where $\dagger$ denotes complex conjugation, while $\mathcal{F}_{\mu \nu}= \partial_{\mu}   \mathcal{A}_{\nu} - \partial_{\nu}   \mathcal{A}_{\mu}$. Expressing $\mathcal{A_{\mu}}$ as $\mathcal{A_{\mu}}=A_{\mu} + i C_{\mu}$, the action~(\ref{dual-symmetric_action complex}) takes the form of the duality-symmetric action~(\ref{dual-symmetric_action}). 

The field equation for the complex potential is
\begin{align}\label{Maxwell eq potential complex}
   \Box \mathcal{A}_{\mu}- \partial_{\mu}\partial^{\nu}\mathcal{A}_{\nu}=0.
\end{align}
This equation is invariant under the infinitesimal conformal transformations $L_{\xi} \mathcal{A}_{\mu} $ in Eq.~(\ref{Lie deriv}) (with $A_{\mu}$ replaced by $\mathcal{A}_{\mu}$), as well as under the ``hidden'' transformations $T_{\xi} \mathcal{A}_{\mu} $ in Eq.~(\ref{extra symmetry})~(with $A_{\mu}$ replaced by $\mathcal{A}_{\mu}$). As in the case of the real potential, the structure of the algebra generated by the conformal and the ``hidden'' transformations is determined by the commutators in Eqs.~(\ref{[Lie,Lie]})-(\ref{[extra,extra]}) with $A_{\mu}$ replaced by $\mathcal{A}_{\mu}$ and $F^{\sigma \rho}$ replaced by $\mathcal{F}^{\sigma \rho}$. 

Now, we will show that if we redefine the ``hidden'' transformations~$T_{\xi} \mathcal{A}_{\mu} $, the ``hidden'' algebra of Eq.~(\ref{Maxwell eq potential complex}) will be isomorphic to $so(4,2) \bigoplus so(4,2)$. Let us redefine $T_{\xi} \mathcal{A}_{\mu} $ by multiplying with $i$ as 
\begin{align}\label{extra symmetry redef complex}
    T'_{\xi}\mathcal{A}_{\mu}\equiv   i\,T_{\xi}\mathcal{A}_{\mu}=&\,~ i\,\xi^{\rho}\epsilon_{\rho \mu \sigma \lambda } \partial^{\sigma} \mathcal{A}^{\lambda},\hspace{5mm} \xi \in so(4,2).
  \end{align}
These transformations leave both the action~(\ref{dual-symmetric_action complex}) and Eq.~(\ref{Maxwell eq potential complex}) invariant~(on the other hand, $T_{\xi}$ is a symmetry of the field equation only). Now, the ``hidden'' algebra of the field equation~(\ref{Maxwell eq potential complex}) is generated by the conformal transformations $L_{\xi}  \mathcal{A}_{\mu}$ and the redefined ``hidden'' transformations $T_{\xi}'  \mathcal{A}_{\mu}$. If we now define the new generators:
\begin{align}
    \mathcal{T}^{\pm}_{\xi} \mathcal{A}_{\mu} 
 \equiv\frac{1}{\sqrt{2}}   \left( L_{\xi}   \pm  T'_{\xi}   \right)\mathcal{A}_{\mu} ,\hspace{5mm} \xi \in so(4,2),
\end{align}
it is easy to see that the $\mathcal{T}^{+}_{\xi}$'s generate a $so(4,2)$ algebra on their own, and so do the transformations $\mathcal{T}^{-}_{\xi}$, while $[\mathcal{T}^{+}_{\xi},\mathcal{T}^{-}_{\xi'} ]=0$ for any $\xi, \xi' \in so(4,2)$~[these follow directly from Eqs.~(\ref{[Lie,Lie]})-(\ref{[extra,extra]})]. Thus, the ``hidden'' algebra is now isomorphic to $so(4,2) \bigoplus so(4,2)$ and closes up to gauge transformations of the complex potential $\mathcal{A}_{\mu}$. 

\section{Zilch continuity equations in the presence of electric charges and currents from symmetries of the standard interacting action}\label{Section_continuity_inter}

In the presence of a non-dynamical material four-current, $J^{\mu}=(\rho,\bm{j})$, the standard interacting EM action is
\begin{align}\label{intercting_action}
    S'=&~ S+ \int d^{4}x\,J^{\nu}A_{\nu}= \int d^{4}x \left(-\frac{1}{4}F^{\mu \nu}F_{\mu \nu}+ 
 J^{\nu}A_{\nu} \right).
\end{align}
Let us consider the variation of $S'$ under the following simultaneous transformations of $A_{\nu}$ and $J^{\nu}$: 
\begin{align}
    \Delta A_{\nu} &= n^{\rho}\tilde{n}^{\mu}\,  \epsilon_{\mu \nu \sigma  \lambda} \,\partial^{\sigma}\partial_{\rho}A^{\lambda} , \label{transform_zilch_vecpot_AND_current_1} \\
    \Delta J^{\nu} &= n^{\rho}\tilde{n}^{\mu}\,  \epsilon_{\mu \hspace{2mm}\sigma  \lambda}^{\hspace{2mm}\nu}\,\partial^{\sigma}\partial_{\rho}J^{\lambda} ,\label{transform_zilch_vecpot_AND_current_2}
\end{align}
where $n^{\rho}$ and $\tilde{n}^{\mu}$ are two arbitrary constant four-vectors, while Eq.~(\ref{transform_zilch_vecpot_AND_current_1}) coincides with the zilch symmetry transformation~(\ref{transform_zilch_vecpot}). The variation of the free action, $S$, is already known to be a total divergence [see Eq.~(\ref{variation_free_action})]. Also, after a straightforward calculation, we find that the variation of the interaction term is a total divergence, as
\begin{align}
    \Delta \Big( \int  d^{4}x \,J^{\nu}A_{\nu} \Big)=&\int  d^{4}x \,\left(\Delta J^{\nu}A_{\nu} + J^{\nu} \Delta A_{\nu} \right)  \\
   =&~\int d^{4}x\,\partial_{\nu}D^{\nu}_{int} \nonumber ,
\end{align}
where
\begin{align}\label{Dint}
   D_{int}^{\nu}= \tilde{n}^{\mu}n^{\rho}(\delta_{\rho}^{\nu}\,J^{\lambda}\,^{\star}F_{\mu \lambda}- \partial_{\rho}J^{\lambda} A^{\alpha}
\epsilon_{\mu \lambda \hspace{2mm} \alpha}^{\hspace{4mm} \nu} ).
\end{align}
Thus, the variation of the interacting action is 
\begin{align}\label{variation_interacting_action}
    \Delta S '= \int d^{4}x \,\partial_{\nu}\left(D^{\nu}+D_{int}^{\nu}   \right),
\end{align}
where $D^{\nu}$ is given by Eq.~(\ref{Dnu}).

Now, by applying the standard Noether algorithm~\cite{weinberg_1995}, we find the following continuity equations for the zilch tensor:
\begin{align}\label{cont_equation_charges_from_Noether}
    \partial_{\lambda}Z^{\mu \lambda\nu}=J_{\lambda}\,\partial^{\nu}\,^{\star}F^{\mu \lambda}-\,^{\star}F^{\mu \lambda} \, \partial^{\nu}J_{\lambda} .
\end{align}
 These  continuity equations determine the rate of gain or loss of the quantity $\int d^{3}x Z^{\mu 0 \nu}$, with spatial density given by $Z^{\mu 0 \nu }$ and flux components given by $Z^{\mu j \nu}$. 

The continuity equations~(\ref{cont_equation_charges_from_Noether}) can be re-expressed in the form of continuity equations for the zilches~(\ref{def_the_zilches}), with spatial density given by $Z^{\mu  \nu 0}$ and flux components given by $Z^{\mu  \nu j}$, as follows. 
First, we observe that, although in the presence of electric charges and currents the quantity $\int d^{3}x Z^{\mu 0 \nu}$ and the $\mu \nu$-zilch, $\int d^{3}x Z^{\mu \nu 0}$~[Eq.~(\ref{def_the_zilches})] are not equal to each other unless $\mu=\nu=0$ (because the symmetry property~(\ref{symmetry_zilches}) no longer holds), they are related to each other by~\footnote{Equation~(\ref{Kibble's_Eq_14}) is obtained following analogous steps as the ones described by Kibble in order to derive equation~(14) (for the free EM field) in Ref.~\cite{Kibble}. Our Eq.~(\ref{Kibble's_Eq_14}) coincides with equation~(14) of Ref.~\cite{Kibble} if no electric charges and currents are present.}
\begin{align}\label{Kibble's_Eq_14}
    Z^{\mu \nu \rho}- Z^{\mu \rho \nu}= \frac{1}{2} \Big(& \epsilon^{\kappa \mu \nu \rho} \,\partial_{\sigma}T^{\sigma}_{\hspace{1mm}\kappa}-\epsilon^{\kappa \lambda \nu \rho} \partial_{\lambda}T^{\mu}_{\hspace{2mm} \kappa}  \nonumber \\
   &-\epsilon^{\kappa \rho \lambda \mu}\partial_{\lambda}T^{\nu}_{\hspace{2mm}\kappa} + \epsilon^{\kappa \nu \lambda \mu}\partial_{\lambda}T^{\rho}_{\hspace{2mm}\kappa} \nonumber\\
   &-2F^{\mu}_{\hspace{2mm}\lambda} \,\epsilon^{\rho \lambda \nu \sigma }J_{\sigma}+2J^{\mu}\,^{\star}F^{\nu \rho}\Big),
\end{align}
where
\begin{align}
    T^{\alpha}_{\hspace{2mm}\beta} =- F^{\alpha \lambda}F_{\lambda \beta}  - \frac{1}{4}\delta^{\alpha}_{\beta} F^{\kappa \lambda}F_{\kappa \lambda}
\end{align}
is the Maxwell stress-energy tensor (with $\partial_{\alpha}T^{\alpha}_{\hspace{2mm}\beta} = J^{\alpha}F_{\alpha \beta}$).
Then, by taking the divergence of Eq.~(\ref{Kibble's_Eq_14}) with respect to the index $\rho$ and using the continuity equation~(\ref{cont_equation_charges_from_Noether}) we find
\begin{align}\label{zilch_cont_charges}
    \partial_{\rho}Z^{\mu \nu \rho}=& ~\eta^{\mu \nu}\,^{\star}F_{\lambda \sigma}\,\partial^{\lambda}J^{\sigma}-\,^{\star}F^{\mu \sigma} \,(\partial^{\nu}J_{\sigma}- \partial_{\sigma}J^{\nu})\nonumber\\
    &-\,^{\star}F^{\nu \sigma} \,(\partial^{\mu}J_{\sigma}- \partial_{\sigma}J^{\mu}).
\end{align}
These are the ten continuity equations determining the rate of gain (or loss) for the ten zilches~(\ref{def_the_zilches}) in the presence of electric charges and currents.
For $\mu=\nu = 0$, both continuity equations~(\ref{cont_equation_charges_from_Noether}) and~(\ref{zilch_cont_charges}) coincide with the known equation~(\ref{conservation optical_sources}) for optical chirality. To the best of our knowledge, the other nine continuity equations in Eq.~(\ref{zilch_cont_charges}) are presented here for the first time. 


\section{{An interesting open question}}\label{Sec_open_question}

Let us suppose that the EM field interacts with a dynamical matter field with corresponding four-current $\tilde{J}^{\mu}$. Now, the action of the full interacting theory is
\begin{align}\label{action_full_dynamical_matter}
    \int d^{4}x \left(   -\frac{1}{4}F_{\mu \nu}F^{\mu \nu} +\tilde{J}^{\nu}A_{\nu}  \right) + S_{\text{matter}},
\end{align}
where $S_{\text{matter}}$ is the action corresponding to the free matter field. According to our earlier discussion, the simultaneous transformations~(\ref{transform_zilch_vecpot_AND_current_1}) and (\ref{transform_zilch_vecpot_AND_current_2}) (with $J^{\nu}$ replaced by $\tilde{J}^{\nu}$) are symmetries of the first two terms in Eq.~(\ref{action_full_dynamical_matter}). Motivated by this observation, we may pose the question of whether one could identify symmetries of the full interacting theory (i.e. symmetries of all three terms in Eq.~(\ref{action_full_dynamical_matter})). In other words, is it possible to identify a transformation of the matter field such that: this transformation is a symmetry of $S_{\text{matter}}$, while the four-current $\tilde{J}^{\mu}$ transforms as in Eq.~(\ref{transform_zilch_vecpot_AND_current_2})? 

\section{{Discussion}}

The results of the present Letter establish a clear connection between all zilch continuity equations and symmetries of the standard EM action via Noether's theorem. Having identified all zilches with Noether charges, we can interpret them as the generators of the corresponding symmetry transformations~(\ref{transform_zilch_vecpot}) of the four-potential in the standard (classical or quantum) EM theory~\cite{weinberg_1995, Generatorsetc, Philbin}. In the case of optical chirality, the explicit knowledge of the underlying symmetry generator is known to offer physical insight, since it allows the identification of the optical chirality eigenstates with plane waves of circular polarization~\cite{Philbin}. Similarly, the symmetry transformations~(\ref{transform_zilch_vecpot}) can be used to identify the eigenstates of all zilches, which is something that we leave for future work.

A particularly interesting uninvestigated question is the one concerning the role of all zilches in light-matter interactions - the case of optical chirality is the only exception since its role has been studied~\cite{optical}. The importance of this question becomes manifest by considering the fact that a physical interpretation for all zilches has been recently provided~\cite{Smith_2018}. In particular, in Ref.~\cite{Smith_2018} it was found that the zilches of a certain class of topologically non-trivial EM fields in vacuum can be expressed in terms of energy, momentum, angular momentum and helicity of the fields. Also, it was demonstrated that the zilches of these fields encode information about the topology of the field lines.
We hope that the results presented in this Letter will be useful in future attempts to study the role of all zilches in light-matter interactions.
More specifically, motivated by the interpretation and applications of the known continuity equation~(\ref{conservation optical_sources}) for optical chirality~\cite{optical,chirality-hel-etc,Poulikakos, optical_helicity_vs_chirality, Generatorsetc}, it is natural to interpret each of our new zilch continuity equations [Eq.~(\ref{zilch_cont_charges})] as determining the rate of loss or gain of the corresponding ``zilch quantity'' of the EM field.

\acknowledgements
The author is grateful to Atsushi Higuchi for useful discussions, encouragement, and comments on earlier versions of this Letter. He also thanks the referee for useful suggestions and comments. Also, it is a pleasure to thank Charis Anastopoulos, Haralampos Geranios, Ed Corrigan, Dmitri Pushkin, Maximilian Ruep, David Serrano Blanco, Nikolaos Koutsonikos-Kouloumpis, Volodimir Simulik and F.F. John for useful discussions. The author acknowledges financial support from the Department of Mathematics, University of York, and from the WW Smith Fund.

\noindent\textbf{Conflict of interest -} On behalf of all authors, the corresponding author states that there is no conflict of interest.

\noindent\textbf{Data availability -} Data sharing is not applicable to this article as no new data were created or analyzed in this study.

\begin{widetext}
\appendix

\section{Invariance of the standard free EM action under the zilch symmetries} \label{appendix_invariance}
Here we present some details for the calculation concerning the invariance of the standard free EM action~(\ref{action}) under the zilch symmetry transformation~(\ref{transform_zilch_vecpot}). For convenience, we focus only on the invariance of the action and not on the derivation of the associated Noether current~(\ref{Noether current zilch}). Also, we drop all terms that are total divergences in order to simplify the presentation. However, note that one needs to keep all such terms if they wish to re-derive Eq.~(\ref{variation_free_action}).

Varying the action~(\ref{action}) with respect to the zilch symmetry transformation~(\ref{transform_zilch_vecpot}) we find
\begin{align}\label{appendix_first}
 -2 \, \Delta S & = \int d^{4}x\, F^{\mu \nu}  \,\Delta F_{\mu \nu}  \nonumber \\
   & =\int d^{4}x\,\left( F^{\mu \nu} \tilde{n}^{\alpha}n^{\rho}\, \partial_{\rho}\partial_{\alpha}\hspace{0.05mm}^{\star}F_{\mu \nu}-F^{\mu \nu} \tilde{n}^{\alpha}n^{\rho}\epsilon_{\alpha \mu \nu \sigma}\partial_{\rho}\partial_{\lambda}F^{\lambda \sigma} \right),
\end{align}
where in the second line we used Eq.~(\ref{transform_zilch_vecpot_FIELDSTRENGTH}). The first term is readily shown to be equal to a total divergence as follows:
\begin{align*}
  \int d^{4}x\, F^{\mu \nu} \tilde{n}^{\alpha}n^{\rho}\, \partial_{\rho}\partial_{\alpha}\hspace{0.05mm}^{\star}F_{\mu \nu} &= -  \int d^{4}x~ \partial_{\rho}F^{\mu \nu} \tilde{n}^{\alpha}n^{\rho}\, \partial_{\alpha}\hspace{0.05mm}^{\star}F_{\mu \nu} \\
  &= 2 \int d^{4}x~ \partial^{\nu}F_{\rho}^{\hspace{1.5mm}\mu }~ \tilde{n}^{\alpha}n^{\rho}\, \partial_{\alpha}\hspace{0.05mm}^{\star}F_{\mu \nu} \\
   &= 2 \int d^{4}x~ \partial^{\nu}\left(F_{\rho}^{\hspace{1.5mm}\mu }~ \tilde{n}^{\alpha}n^{\rho}\, \partial_{\alpha}\hspace{0.05mm}^{\star}F_{\mu \nu}\right),
\end{align*}
where in the second line we used Eq.~(\ref{Bianchi_usual}), and in the third line we used that the divergence of $^{\star}F_{\mu \nu}$ vanishes identically because of Eq.~(\ref{Bianchi_usual}). We now drop the first term in Eq.~(\ref{appendix_first}) (since we showed that it is a total divergence) and we express Eq.~(\ref{appendix_first}) as
\begin{align}
  -2\, \Delta S &  =-2\int d^{4}x~ \hspace{0.05mm}^{\star}F_{\alpha \sigma}~\tilde{n}^{\alpha}n^{\rho}\,\,~\partial_{\rho}\partial_{\lambda}F^{\lambda \sigma} .\label{OK_mate}
\end{align}

On the other hand, keeping both terms in Eq.~(\ref{appendix_first}) and using the off-shell identity~(\ref{Bianchi dual}), Eq.~(\ref{appendix_first}) is re-written as
\begin{align}\label{appendix_second}
  -2\, \Delta S & =2\int d^{4}x~ F^{\nu \sigma}\tilde{n}^{\alpha}n^{\rho} \, \,\partial_{\rho}  \partial_{\nu}\hspace{0.05mm}^{\star}F_{\alpha  \sigma}.
\end{align}
Integrating by parts twice, we find
\begin{align}
  -2\, \Delta S & =2\int d^{4}x~ \hspace{0.05mm}^{\star}F_{\alpha \sigma}~\tilde{n}^{\alpha}n^{\rho}\,\,~\partial_{\rho}\partial_{\lambda}F^{\lambda \sigma}  .\label{appendix_fifth}
\end{align}
Comparing this equation with Eq.~(\ref{OK_mate}), we find $\Delta S = 0$ (i.e. $\Delta S$ is equal to the integral of a total divergence), as required.

\end{widetext}

\nocite{*}


\providecommand{\noopsort}[1]{}\providecommand{\singleletter}[1]{#1}%

\end{document}